\documentstyle{aipproc}
\input epsf       

\begin{document}

\centerline{{\it Proceedings of the 5$^{th}$ Huntsville Gamma-Ray Bursts Symposium (1999)}}
\vspace{0.5cm}

\title{BeppoSAX and Ulysses data on the giant flare from SGR 1900+14}

\author{M.~Feroci$^{(1)}$ K.~Hurley$^{(2)}$ R.~Duncan$^{(3)}$
C.~Thompson$^{(4)}$ \\
E.~Costa$^{(1)}$ and F.~Frontera$^{(5),(6)}$}
\address{$^{(1)}$Istituto di Astrofisica Spaziale - CNR, Via del Fosso del Cavaliere, 00133 Roma, Italy\\
$^{(2})$University of California, Space Sciences Laboratory,
Berkeley, CA 94720-7450\\
$^{(3)}$University of Texas, Department of Astronomy, Austin,
TX 78712, USA\\
$^{(4)}$University of North Carolina, Department of Physics and Astronomy,
Philips Hall, Chapel Hill, NC 27599-3255\\
$^{(5)}$TESRE/CNR, Via Gobetti 101, Bologna, Italy\\
$^{(6)}$Dipartimento di Fisica, Universit\'a di Ferrara, Via del Paradiso 12, Ferrara, Italy}

%\lefthead{LEFT head}
%\righthead{RIGHT head}
\maketitle

\begin{abstract}
The extraordinary giant flare of 1998 August 27 from 
SGR 1900+14 was the most intense event 
ever detected from this or any other cosmic source (even more intense
than the famous March 5th 1979 event).
It was longer than any previous burst from
SGR1900+14 by more than one order of magnitude, and
it displayed the same 5.16-s periodicity in hard X-rays that was detected
in the low energy X-ray flux of its quiescent counterpart.
The event was detected by several gamma-ray
experiments in space, among them the Ulysses gamma-ray burst detector
and the BeppoSAX Gamma Ray Burst Monitor. These instruments operate
in different energy ranges, and a comparison of their data shows
that the event emitted a strongly energy-dependent flux, and displayed
strong spectral evolution during the outburst itself.
Here we present a joint analysis of the BeppoSAX and
Ulysses data, in order to identify the energy-dependent features
of this event and understand some of the physical conditions
in the environment of the neutron star which generated this flare.
\end{abstract}

\section*{Introduction}

After several years of quiescence SGR 1900+14 became 
extremely active in May 1998.
ASCA and RXTE observations after the 1998 May activity episode
revealed a periodicity of 5.16~s in the quiescent
X--ray (2--10 and 2--20 keV, respectively) emission  with period derivative
$\sim 1 \times 10^{-10} \rm  s \, s^{-1}$
\cite{hurley99a,kouveliotou99}.  
This spindown rate is consistent with a magnetar-strength field
\cite{duncan92,thompson95}.

On 1998 August 27 a giant flare from SGR1900+14, lasting more than
five minutes, was detected by Konus-Wind, \it Ulysses \rm  BeppoSAX
and NEAR \cite{hurley99b,feroci99,mazets99}.
Gamma rays during the first second were extraordinarily intense, overwhelming
detectors on several other spacecraft as well.
The Compton Gamma-Ray Observatory was Earth-occulted for this flare.
The 5.16~s neutron star rotation period
was strongly detected during the
giant flare.  Indeed, the periodic signal was intense enough to produce a
marked 5.16-second modulation in the height of the Earth's ionosphere, which
affected long-wavelength radio transmissions \cite{inan99}.
After $\sim40$ seconds, the flare
evinced a $\sim$1.03~s repetitive pattern that is unlike any emission
previously detected from any source \cite{feroci99,mazets99}.
A radio afterglow was found with the Very Large Array
\cite{frail99}.  This source
was apparent in the error box of SGR1900+14 on 1998 September 3,
but it faded away in less than one week, providing evidence for an
abrupt outflow of relativistic particles during the flare.

In this paper we show preliminary results of a comparative analysis of 
BeppoSAX and {\it Ulysses} observations of the August 27th event.
Additional analysis, results and their interpretation 
may be found in \cite{feroci00}.

\section*{The instruments}

The BeppoSAX Gamma Ray Burst Monitor (GRBM, \cite{frontera97,feroci97}) 
consists of the four anticoincidence CsI(Na) detectors of the
Phoswich Detection System (PDS, \cite{frontera97}),
forming a square box, surrounding the main PDS detectors and
located in the core of the BeppoSAX payload.
Each shield is 1 cm thick and has dimensions 27.5 x 41.3 cm. 
The GRBM electronics records data from each shield with both low (1~s)
and high ($<$8~ms) time resolution. 
The 1~s data consist of count rates in the 40-700 and $>$100 keV energy
ranges. Additional details on the GRBM may be found in \cite{feroci97}.

The \it Ulysses \rm GRB detector \cite{hurley92}
consists of two 3 mm thick
hemispherical CsI(Na) scintillators
with a projected area of about 20 cm$^2$
in any direction.  The detector is mounted
on a magnetometer boom far from the body of
the spacecraft.
The energy range is $\sim$ 25-150 keV.
The lower energy threshold is set by a discriminator,
and is in practice an approximate
one; photons with energies $>10$ keV can penetrate
the housing and be counted either
because of the rather poor energy resolution at
low energies, or, in the case of
very intense events, due to pulse pile-up.
The instrument takes time history data with time resolutions
of 31.25 ms for 64 s, and 0.5 s for the full duration of the event.

\section*{Data Analysis}

\subsection*{Low Time Resolution Data}

\begin{figure}
\epsfxsize=10.0cm \centerline{\epsfbox{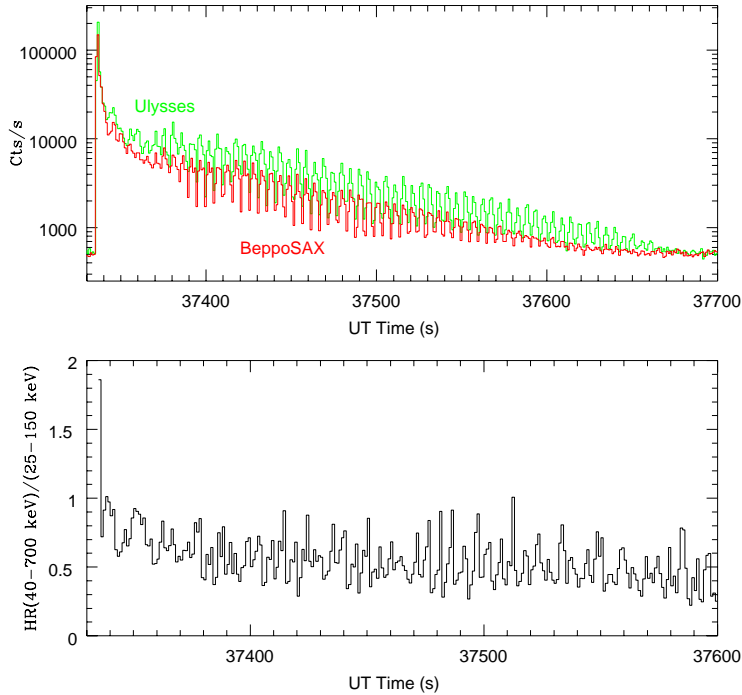}}
\caption{Top panel: BeppoSAX/GRBM (40-700~keV)
and \it Ulysses\rm' GRB detector (25-150~keV) 1-s light curve of the
giant ourburst. A constant value of 500 cts/s has been added to the
background-subtracted GRBM data for display purposes. 
Time reference is seconds of day August 27 1998.
Bottom panel: Hardness Ratio between the light curves in the two 
energy ranges (typical 1-$\sigma$ error $\sim$0.01-0.02).
The \it Ulysses \rm time has been shifted to account for its 
position with respect to the Earth.}
\label{lrlc}
\end{figure}

In Figure~\ref{lrlc} (top panel) we show the event light curve
as detected by the two above experiments.
The 5.16-s pulsation is evident in both
energy ranges for the entire duration of the event ($\geq$300~s). 
Even if the low
resolution data are independently synchronized with the
onboard clocks of the two spacecrafts, and the event was detected at
different UT times due to spatial separation, the relative timing
between the two light curves turned out to be, purely by chance,
synchronized within approximately 100~ms. This allows one to use 
the light curves in the two energy ranges (25-150 and 40-700~keV)
to produce a hardness ratio as a function of time, with a
time resolution of 1~s. The result is shown in Figure~\ref{lrlc} 
(bottom panel).

\begin{figure}
\epsfxsize=10.0cm
\epsfysize=6.0cm
%\centerline{\epsfbox{fig.ps}}
\centerline{\epsfbox{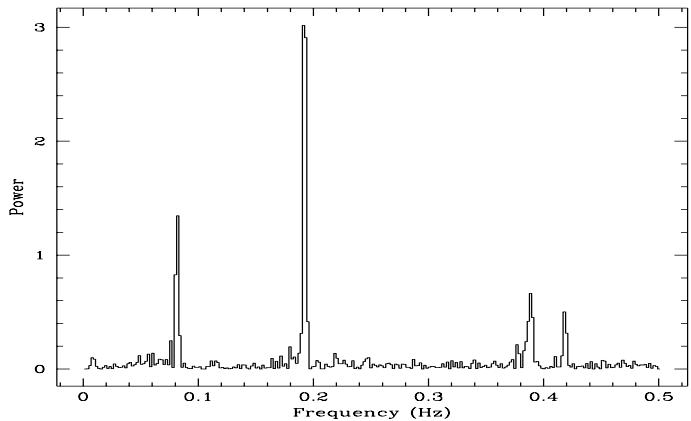}}
\caption{Power density spectrum of the hardness ratio curve shown in
the bottom panel of Figure~\ref{lrlc}}
\label{lr_psd}
\end{figure}

It is clear from the hardness ratio plot that the energy spectrum
of the emitted radiation significantly changes during 
each 5.16-s pulse. In fact,
performing a Fast Fourier Transform of the hardness ratio
curve we obtain the power density spectrum presented in
Figure~\ref{lr_psd}, where the fundamental harmonics at
$\sim$0.2~Hz is clearly detected.
The minor peaks are easily identified as well. The two peaks
around 0.4~Hz derive from aliasing of the higher order
harmonics, due to the limited frequency span.
The peak at $\sim$0.08~Hz is actually spurious, deriving from
the spinning motion of the \it Ulysses \rm spacecraft
at about 5 rpm. The modulation in the \it Ulysses \rm count rate
is likely due to a partial occultation by the carbon fiber
magnetometer boom.

\subsection*{High Time Resolution Data}

\begin{figure}
\epsfxsize=9.0cm \centerline{\epsfbox{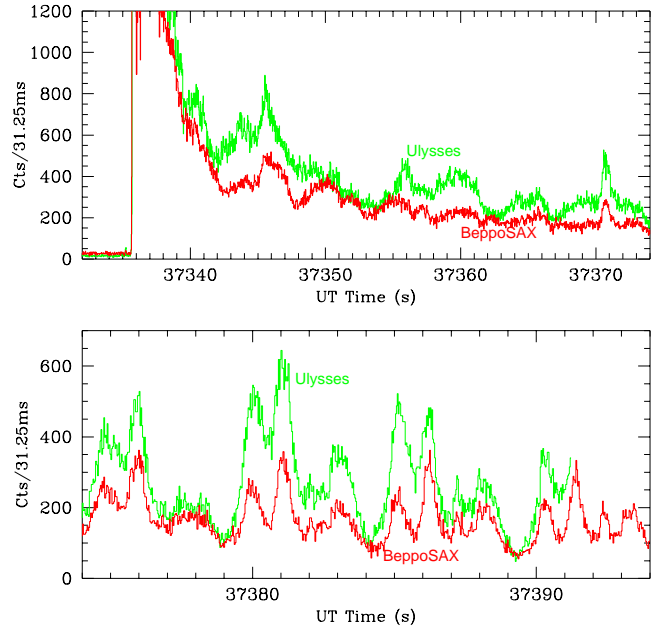}}
\caption{Hight time resolution (31.25~ms) light curves of
an initial $\sim$60~s portion of the giant outburst 
from the \it Ulysses \rm (25-150~keV)
and BeppoSAX (40-700~keV) GRB experiments. The \it Ulysses \rm
time has been shifted to account for its position with
respect to the Earth.}
\label{hrlc}
\end{figure}

High time resolution data are available from both the instruments
for a limited lapse of time. In particular, these \it Ulysses \rm
data stop about 60~s after the event onset, whereas the BeppoSAX/GRBM
recorded the high resolution light curve for approximately
106~s after the trigger.
In Figure~\ref{hrlc} they are shown with a time resolution of 31.25~ms,
over the time interval in which they are both available.

\section*{Conclusions}

The data from experiments operating at different energy ranges
(the \it Ulysses \rm count rate is dominated by $\sim$10-40~keV 
photons, whereas
only photons above $\sim$40~keV contribute to the BeppoSAX count rate) 
allow for a time-resolved study of the giant outburst from SGR 1900+14.

From the low time resolution data it appears that the energy spectrum
of the emitted X-ray flux is strongly modulated by the 5.16-s rotation
of the neutron star.
From the high time resolution data in Figure~\ref{hrlc}, 
we see that the relative contribution of soft (i.e., \it Ulysses \rm)
and hard (i.e., GRBM) X-rays changes over the spin phase,
in a different way in subsequent pulses, at least during the initial 
$\sim$60~s of the event.
It is also interesting to note that the soft  
counts usually exceed numerically the hard ones, 
except near the 5.16-s pulse minima, where they are basically equal. This
indicates that harder emission persists during the occultation of a softer
radiation beam.

Hence, both data sets indicate significant spectral
evolution over the spin phase and from one pulse to another. 
These data likely reflect the complex
behaviour of bubbles of relativistic plasma trapped by the star's 
magnetosphere following a transient outflow that gave rise to the
hard initial spike in the event light curve. 
The late-time stability of the four-peaked pulse shape (see also
\cite{feroci99,mazets99}) strongly suggests a multipolar structure 
for the magnetic field.
These facts, in addition to the harder emission at the minima, 
suggest the existence of two spatially distinct emission components
dominating the soft and hard energy domains, 
undergoing separate histories both in time and space.


\begin{references}

\bibitem{hurley99a}
Hurley, K. et~al. 1999a, ApJ 510, L111

\bibitem{kouveliotou99}
Kouveliotou, C. et~al. 1999, ApJ 510, L115

\bibitem{duncan92}
Duncan, R.C. \& Thompson, C., 1992, ApJ, 392, L9

\bibitem{thompson95}
Thompson, C., and Duncan, R.C., 1995, MNRAS, 275, 255

\bibitem{hurley99b}
Hurley, K. et~al. 1999b, Nature 397, 41

\bibitem{feroci99}
Feroci, M. et~al. 1999, ApJ 515, L9

\bibitem{mazets99}
Mazets, E.P., et al., 1999, preprint (astro-ph/9905196 v2)

\bibitem{inan99}
Inan, U., et~al. 1999, Geophys. Res. Lett., 26(22), 3357

\bibitem{frail99}
Frail, D., Kulkarni, S., and Bloom, J., 1999, Nature 398, 127

\bibitem{feroci00}
Feroci, M. et~al. 1999, \it in preparation \rm

\bibitem{frontera97}Frontera, F. et. al. 1997, Astron. Astrophys. Suppl. Ser., 
122, 357

\bibitem{feroci97}
Feroci, M. et~al. 1997, SPIE Proceedings, 3114, 186

\bibitem{hurley92}
Hurley, K., et al. 1992, Astron. Astrophys. Suppl. Ser., 92(2), 401



\end{references}
\end{document}